\documentclass{article}
\usepackage{graphicx}
\renewcommand{\d}{\partial}
\makeatletter
\renewcommand{\section}{\@startsection{section}{1}{0pt}%
{2ex plus 0.5ex minus 0.2ex}{0.2cm}{\large\sc}}
\renewcommand{\subsection}{\@startsection{subsection}{2}{\parindent}{0.3cm}{-1em}{\normalsize\bf}}
\makeatother
\begin{document}
\twocolumn
[
\begin{center}
{ \Large \bf
Absorption of dark matter by a supermassive black hole\\ at the Galactic center: 
role of boundary conditions

\medskip}
{ \large \bf
M.I.\,Zelnikov, E.A.\,Vasiliev

\medskip}
{\em
Lebedev Physical Institute, Moscow, Russia\\
e-mail: zelnikov@lpi.ru, eugvas@lpi.ru}

\smallskip
{JETP Letters, Vol.81, No.3, 2005, pp.85-89

Received December 15, 2004}

\begin{quote}
The evolution of the dark matter distribution at the Galactic center is 
analyzed, which is caused by the combination of gravitational scattering on 
Galactic bulge stars and absorption by a supermassive black hole at the 
center of the bulge. Attention is focused on the boundary condition on the 
black hole. It is shown that its form depends on the energy of dark matter 
particles. The modified flux of dark matter particles onto the black hole is 
calculated. Estimates of the amount of dark matter absorbed show that the 
fraction of dark matter in the total mass of the black hole may be significant. 
The density of dark matter at the central part of the bulge is calculated. It 
is shown that recently observed $\gamma$ radiation from the Galactic center can be 
attributed to the annihilation of dark matter with this density.

\end{quote}
\end{center}

]{
\section{Introduction}
Investigations of the distribution of dark matter in the nuclei of galaxies 
currently attract the interest of many researchers in view of both the effect 
of dark matter on the growth of supermassive black holes and the search for 
its possible annihilation radiation. In this work, we analyze the evolution of 
the dark matter distribution at the Galactic center under the effects of 
scattering by stars at the Galactic center and absorption by the central black 
hole. We assume that dark matter consists of particles that undergo only 
gravitational interaction and possibly slightly annihilate (so-called weakly 
interacting massive particles, WIMPs) \cite{DMreview}.

The interaction of the central black hole with its environment, including dark 
matter at the Galactic center (bulge), was analyzed in \cite{GS, Merritt, IGZ, 
SIGZ, ZVl} and elsewhere. An 
approach in which the distribution function of the dark matter is written in 
terms of invariants of motion seems to be most consistent. In the case of 
spherically symmetric, sufficiently slow evolution (compared to dynamical 
time), such variables are the orbital angular momentum $m$, its projection $m_z$ 
onto the $z$ axis, and the radial action (adiabatic invariant)
$$
I=\frac{1}{\pi}\int_{r_-}^{r_+}v(r)\,dr \,.
$$
We use a model of the dark matter halo formation in which its spatial 
density profile before the formation of the Galaxy and baryonic compression 
had the form
$\rho(r) \sim r^{-12/7}$, and the initial distribution function in terms $I, m, 
m_z$ has the form \cite{GZ}
\begin{equation}  \label {f_init}
f(I, m, m_z) = f_0\,{I}^{1/8}\,\delta(m^2-{l_0}^2 {I}^2)\,,
\end{equation}
where $l_0 \sim 0.1$ is a small parameter.
We emphasize the convenience of choosing radial action $I$ rather than particle
energy $E$ as a basic variable when analyzing evolution in a slowly varying 
potential.

\subsection{Kinetic equation.}
In the zeroth approximation, the distribution function of dark matter in the 
chosen variables does not change upon baryonic compression. Under the 
assumption that the black hole is formed at the coordinate origin, the increase 
in the black hole mass $M_{bh}$ causes the absorption of particles whose angular 
momenta $m$ are less than the critical value $m_g = 4\,G\,M_{bh}/c$, 
but the amount of such particles turns out to be negligibly small 
\cite{IGZ}. However, the situation is 
significantly changed due to effects associated with the gravitational 
scattering of dark matter particles on bulge stars. These effects lead to the 
diffusion of particles in phase space \cite{IGZ, ZVl}.

In the first approximation, the evolution of the distribution function is 
described by the kinetic equation
\begin{equation}  \label{kineq0}
\frac{\d f(\{I_i\},t)}{\d t} = \frac{\d}{\d I_k} \,,
  \left[R_{kl}\frac{\d f}{\d I_l}\right]
\end{equation}
where $\{I_i\} = \{I, m, m_z\}$, and $R_{kl}$ are the corresponding diffusion 
coefficients. Owing to the spherical symmetry of the evolution, $m_z$ does not 
enter into the equations. In addition, $I$ can be approximated as 
$I(E,m) \approx J(E) - \beta m$ \cite{GS}, 
where $\beta=1$ for the Coulomb potential of the black hole, and 
$\beta \simeq 0.6$ for the isothermal potential of the bulge. 
Hereafter, we will use $J$ instead of $I$, because we are interested in the 
region of low angular momenta. Since the parameter $l_0$ is small, $I$ can also 
be changed to $J$ in the initial distribution (\ref{f_init}).
It can be shown that the cross terms $R_{12}$ are small in the region of 
interest \cite{ZV}. 
Finally, the kinetic equation takes the form
\begin{equation}  \label{kineq}
\frac{\d f(J,m,t)}{\d t} =
  \frac \d{\d J} \left(R_{11}\frac{\d f}{\d J}\right) +
  \frac 1 m \frac \d{\d m}\left(m\,R_{22}\frac{\d f}{\d m}\right) \,.
\end{equation}
The diffusion coefficients are calculated in \cite{ZV}. 
One-dimensional diffusion along the $m$ axis turns out to be the most significant. It 
leads to the particle flux into the region of low orbital angular momenta and 
to their absorption by the black hole for $m \le m_g$. This process was analyzed in 
detail in \cite{IGZ,ZV}. The effect considered there is the following. 
Let us write Eq. (\ref{kineq}) in the one-dimensional form
\begin{equation}  \label{diff1d}
\frac{\d f(J,m,t)}{\d t} = \frac 1 m \frac \d{\d m}
\left(m\,R_{22}\frac{\d f}{\d m}\right) \,.
\end{equation}
The initial condition is taken in the form (\ref{f_init}), and the boundary 
condition at the black hole corresponds to absorption: $f(J,m=m_g,t)=0$. 
The diffusion coefficient $R_{22}$ calculated for the isothermal bulge has the 
form
\begin{equation}  \label{R22b}
R_{22} = 0.46\,G\,M_s\,L_c\,\sigma_0 \,,
\end{equation}
where $\sigma_0$ is the velocity dispersion of bulge stars, $M_s$
is the mass of a star (for simplicity, we assume that all stars have the same 
mass equal to $M_\odot$), and $L_c \simeq 10$ is the Coulomb logarithm.

In this formulation, the solution of the diffusion equation yields the 
following expression for the dark matter flux onto the black hole
\begin{equation}  \label{S_ot_t}
S(t) = 2(2\pi)^3 \int_0^\infty f_0 J^{1/8}\,S_J(t) \propto {R_{22}}^{9/16}\,
t^{-7/16}\,.
\end{equation}
Here
\begin{equation}  \label{SJ_ot_t}
S_J(t) = \frac{0.18}{\ln\frac{l_0\,J}{2m_g}} \cdot \frac{1}{t} 
\exp\left( - \frac{{l_0}^2\,J^2}{5\,R_{22}\,t}\right)
\end{equation}
is the flux in one-dimensional diffusion equation (\ref{diff1d}) with the 
initial condition $f(t=0)=\delta(m^2-(l_0 J)^2)$.

However, the absorbing boundary condition is valid only for $\overline{\Delta
m^2} \ll {m_g}^2$, where
\begin{equation}  \label{Delta_m}
\Delta m = \sqrt{2\,T(J)\,R_{22}} \simeq \frac{\sqrt{2\pi\,R_{22}\,J}}{\sigma_0}
\end{equation}
is the rms change in the orbital angular momentum per orbital period $T(J)$. 
It is easy to show that this condition is violated even at rather low energies 
of the particle. In this work, we obtain a more accurate expression for the 
boundary condition and investigate its effect on the absorption rate of the 
dark matter. We start with the presentation of the Milky Way bulge model.

\subsection{Bulge model.}
We assume that the star distribution in the bulge is isothermal, i.e., 
spherically symmetric and isotropic, and has the power-law density profile
\begin{equation}  \label{nstarbulge}
\rho_{s(out)}(r) = \rho_0 \left(\frac{r}{R_0}\right)^{-2} \;,\qquad
  \rho_0{R_0}^2 = \frac{{\sigma_0}^2}{2\pi\,G} \,,\\
\end{equation}
where $\sigma_0$ is the star velocity dispersion independent of the distance to 
the center. In the inner part of the bulge -- the black hole influence region 
-- the density profile has lower exponent \cite{genzel5}:
\begin{equation}  \label{nstarcentre}
\rho_{s(in)}(r) = \rho_0 \left(\frac{r}{R_0}\right)^{-3/2} \,,
\end{equation}
where $\rho_0$ and $R_0$ are the same values as in (\ref{nstarbulge}), which
assures the continuity of the star density at the boundary of the black hole 
influence region. We define the radius of the influence region such that the 
total mass of stars inside the region is equal to the mass of the black hole:
$$
M_{in} = \int_0^{R_0} \rho_{s,in}(r')\,4\pi r'^2\,dr' = 
  \frac{8\pi}{3}\rho_0{R_0}^{3/2} = M_{bh}\:,\quad
$$
\begin{equation}  \label{sigma_R_0}
\frac{G\,M_{bh}}{R_0} = \frac{4}{3}{\sigma_0}^2
\end{equation}
In Milky Way, $\sigma_0 = 85\div 90$~km/s, and the value of $R_0$ corresponding 
to the observed black hole mass $M_{bh}=3\cdot 10^6 M_\odot$, is $R_0 = 1.3$~pc.
These values also give the observation-consistent normalization for the star 
density profile at the center [See Eq.(\ref{nstarcentre})].

\section{Absorption of particles by the black hole and boundary conditions}
Thus, as was mentioned above, the absorbing-boun\-da\-ry approximation $f(m_g,J,t)=0$
is valid only if $\Delta m \ll m_g$, i.e., change in the orbital angular 
momentum of the particle per period is small compared to the characteristic 
problem scale -- the boundary orbital angular momentum. At the same time, as is 
easily seen from Eqs. (\ref{Delta_m}, \ref{SJ_ot_t}), this condition is violated
for Milky Way at present, because the maximum of the flux comes from the values 
of $J$ for which $\Delta m \gg m_g$.

To describe correctly the absorption of particles by the black hole, we 
consider two limiting cases (see fig-ure): the random-walk approximation 
(absorption for $\Delta m \ll m_g$, and the pinhole approximation for 
$\Delta m \gg m_g$ (the names are taken from Lightman\&Shapiro \cite{LS}, 
where this process was considered in application to stars).

\begin{figure}[h]
$$\includegraphics{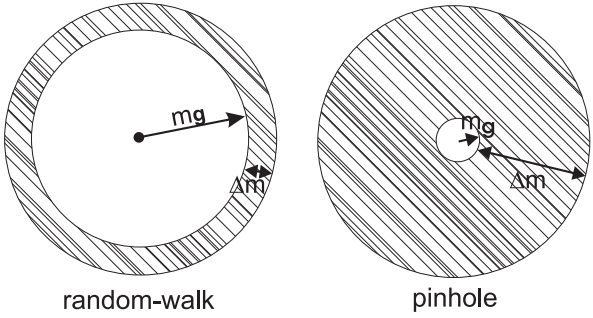} $$
\caption{Two absorption regimes}
\end{figure}

The absorption models are different in the random-walk and pinhole 
approximations. In the former case, the distribution function
$f(m,J,t) \to 0$ for $m\to m_g$, and its derivative at the absorption boundary
$m=m_g$ determines the flux. In the latter case, $f(m)\to f_g$ where $f_g$ is 
a nonzero boundary limit, and the absorption rate is determined as follows. 
There is a nonzero probability per orbital period that the angular momentum 
$m$ of a particle increases by $\delta m$ such that $m+\delta m < m_g$
and, hence, the particle is captured by the black hole. Flux onto the black 
hole is determined as the sum of the absorption probabilities for all particles 
per period. At the same time, it is equal to the diffusion flux from higher $m$ 
values, where the diffusion approximation is valid ($\Delta m < m$).

\subsection{Boundary condition.}

Different absorption\\
 regimes can be described by the following modification of 
the boundary condition: 
\begin{equation}  \label{BCnew}
\left.\left(f - m_g \alpha\frac{\d f}{\d m}\right)\right|_{m=m_g} = 0 \,.
\end{equation}
Values $\alpha \ll 1$ correspond to the random-walk approximation, and
$\alpha \gg 1$ to the pinhole approximation.

Therefore, if we take the value of the distribution function at the boundary 
$f(m=m_g)=f_g$ and assume that the flux is continuous near the boundary and 
varies only slightly at low $m$, then we obtain the following expression for 
$f$ near the boundary:
\begin{equation}  \label{f_near_mg}
f(m) = f_g ( 1 + \frac{1}{\alpha} \ln\frac{m}{m_g} )
\end{equation}
Our aim is to obtain expressions for $\alpha$ and for modified flux $S_J(t)$ of 
particles with radial action $J$ on the black hole in the limit $\alpha \gg 1$.

For this purpose, we calculate the number of particles that diffuse into the 
region $m<m_g$ and are absorbed by the black hole during one orbital period.
We assume that the probability distribution for obtaining a given increment
$\delta m$ of the particle angular momentum is Gaussian with the variance  
$\overline{\delta m^2}=\Delta m^2$ and mean value
$\overline{\delta m}\ll \Delta m$ \cite{LS}:
$$
p(m,\delta m) = \frac{1}{\sqrt{2\pi}\,\Delta m} \exp\left(-\frac{\delta m^2}{2\,
\Delta m^2}\right) \times
$$
\begin{equation}  \label{p_deltam}
\times \frac{1+\delta m/m}{\frac{1}{2}\left[1+{\rm erf}
(\frac{m}{\sqrt{2}\,\Delta m})\right] + \frac{\Delta m}{\sqrt{2\pi}}
\exp(-\frac{m^2}{2\,\Delta m^2})}
\end{equation}
The normalization is taken such that 
$\int_{-m}^\infty p(m,\delta m)\,d\delta m = 1$.

The total number of particles absorbed per period is given by the expression
\begin{equation}  \label{SJT}
Q = \int_{m_g}^\infty 2m\,dm\,f(m)\int_{-m}^{-(m-m_g)} p(\delta m)\,d\delta m
= S_J\,T(J) \,.
\end{equation}
At the same time, it is equal to the flux from larger $m$ values per period:
\begin{equation}  \label{SJT2}
Q = S_J\,T(J) = m\,R\,\frac{\d f}{\d m}\,T = \frac{f_g}{\alpha}\,R\,T =
\frac{f_g\,\Delta m^2}{2\alpha} \,.
\end{equation}
In the limit $\Delta m \gg m_g$, we obtain
\begin{equation}  \label{alpha}
\alpha = 2.8\left(\frac{\Delta m}{m_g}\right)^2 \,,
\end{equation}
which agrees with the result obtained in \cite{LS}.

\subsection{Dark matter absorption.}
To determine the expression for the particle flux onto the black hole 
$S_J(t)$, we use the same procedure as for the pinhole limit considered in
\cite{ZVl}. Namely, knowing the general form of the expression
$S_J(t)\propto\frac{1}{t}\exp(-\frac{{m_0}^2}{5\,R\,t})$,
we combine the solution $f(m,t)$ given by Eq. (\ref{f_near_mg}) and the 
Gaussian diffusion solution for the initial condition
$f(t=0) = \delta(m^2-{m_0}^2)$. Thus, we obtain
\begin{equation}  \label{S_Jt}
S_J(t) = \frac{0.18}{\alpha+\ln\frac{m_0}{2\,m_g}}\,\frac{1}{t}
  \exp\left(-\frac{{l_0}^2\,J^2}{5\,R\,t}\right)
\end{equation}

This expression for flux differs from the one obtained in previous 
investigation (\ref{SJ_ot_t}) by an additional term $\alpha$ in the 
denominator. As was pointed out, at present $\alpha(J)\gg 1$ for the values of 
$J$ from which flux is maximal and, therefore, the total flux is much lower than 
the value obtained without corrections. Indeed, the total flux is given by the 
expression
\begin{eqnarray}  \label{Stotal}
S(t) = 2(2\pi)^3 \int_0^\infty dJ\,f_0\,J^{1/8}\,S_J(t) \approx \nonumber\\
\approx 9.5\,f_0\left(\frac{5\,R_{22}\,t}{{l_0}^2}\right)^{9/16} t^{-1}
  (\beta+1)^{-8/9} 
\end{eqnarray}
where the quantity 
\begin{equation}  \label{beta}
\beta = \frac{\sqrt{5\,R_{22}\,t}\,2\pi R_{22}}{20 l_0 {\sigma_0}^2{m_g}^2}
  \approx 7 \left(\frac{t}{10^{10}\mbox{ лет}}\right)^{\textstyle\frac 1 2}
  \left(\frac{M_{bh}}{3\cdot 10^6 M_\odot}\right)^{-2}
\end{equation}
reflects the correction due to the modification of boundary condition (\ref{BCnew}). 
If the correction is disregarded (by setting $\beta=0$), the expression 
(\ref{Stotal}) yields a power-law increase of the black hole mass
$M_{bh}(t)\propto t^{9/16}$ \cite{ZVl}. 
The estimate for the mass directly depends on the parameters $f_0$ and $l_0$
of the initial dark matter distribution. Taking $f_0=6\cdot 10^9$ 
g (cm$^2$/s)$^{-9/8}$ and $l_0=0.1$ for Milky Way according to \cite{ZVl}, 
we estimate the mass as $M_{bh} \simeq 11\cdot 10^6 M_\odot$ for the present time
($t=10^{10}\mbox{ yr}=3\cdot 10^{17}$~s)
and for negligibly low seed black hole mass (under the assumption that 
mass increases only due to the absorption of dark matter; i.e., it is a 
lower estimate for the mass).

The inclusion of the flux modification significantly changes the situation. As 
follows from Eq. (\ref{beta}), if the black hole mass is low, then $\beta\gg 1$ 
and hence the dark matter flux is also low. This is an evidence of the baryonic 
nature of a seed black hole. An upper estimate for the absorbed dark matter 
mass $M_d$ can be obtained as the difference $M_d = M_0-M_{bh}(0)$ between the 
final mass $M_{bh}(t_0=3\cdot 10^{17}\,{\rm s}) = M_0 = 3\cdot 10^6 M_\odot$, 
and the initial mass $M_{bh}(0)$ obtained by solving the equation 
${dM_{bh}}/{dt} = S(t)$ backwards in time.
For the values $f_0$ and $l_0$ adopted above, we obtain $M_d = 0.67\,M_0 = 2\cdot
10^6 M_\odot$. If the value $f_0$ is ten times lower, then $M_d = 0.11\,M_0$.
Another estimate can be obtained by the supposition that the growth of the 
black hole due to absorption of both dark and baryonic matter has a power-law 
form $M_{bh}(t) = M_0 (t/t_0)^\gamma$. In this case, the mass of dark matter 
absorbed is given by the expression $M_d = \int_0^{t_0} S(t, M_{bh}(t))\,dt$. 
For the growth law with the exponent $\gamma=1/2$ (obtained in \cite{SIGZ}
analyzing the absorption of stars) $M_d=0.36\,M_0$; it is proportional to
the normalization constant $f_0$.

Thus, the correct formulation of the boundary condition considerably 
(approximately by an order of magnitude) reduces the estimate for the mass of 
dark matter absorbed by the black hole. The absorbed mass comprises up to 
$1/3 - 1/2$ of the current black hole mass for the chosen normalization 
of the dark matter density. We emphasize that a similar consideration for 
the absorption of stars shows that the modification of the boundary condition 
is insignificant \cite{SIGZ}. This difference results from the fact 
that dark matter particles initially have low orbital angular momenta and 
sooner reach the absorption boundary. This means that the flux at a given time 
comes from higher $J$ values than those for stars. Hence, the boundary 
condition for dark matter changes its form at large $J$ values more considerably.

\section{Density profile of dark matter and its annihilation}
We consider the detection of $\gamma$ radiation from the Galactic center, which is 
probably due to annihilation of weakly interacting particles of dark matter 
(WIMPs), as a possible observational test of the current dark matter 
distribution. Photons of energies of several TeVs have been detected on the 
H.E.S.S. telescope, whose angular resolution of $3'-5.8'$ corresponds to a spatial
region of about 10 pc in size (the distance to the Galactic center is taken to 
be 8.5 kpc). As was shown in \cite{Horns}, the observed photon flux can be explained by 
the annihilation of supersymmetric particles with a mass of about 12 TeV whose 
density has a power-law profile and a mean value of about
$\sim 10^3 M_\odot/\mbox{pc}^3$ inside the central 10 pc.

Knowing the distribution function of dark matter in phase space, it is easy to 
calculate its density profile as a function of the distance from the coordinate 
origin:
\begin{equation}  \label{rho_r}
\rho(r) = \frac{2\pi}{r^2} \int\limits_{J_{min}}^\infty dJ \frac{\d E}{\d J}
  \int\limits_{m_g}^{m_{max}} 2m\,dm \frac{f(J,m,t)}{\sqrt{2(E-\Psi(r))-\frac{m^2}{r^2}}}
\end{equation}

Let us calculate the dark matter density in the central region 10 pc in size. 
For simplicity, we assume that the potential in this region is determined by 
the central star cluster with a density 
$\rho_s(r) = {\sigma_0}^2/2\pi Gr^2$, where $\sigma_0$ 
is the star velocity dispersion. As follows from (\ref{f_near_mg}, \ref{S_Jt}),
$$
f(J,m,t) = f_0J^{1/8}\,\frac{S_J(t)}{R} \left(\alpha + \ln \frac{m}{m_g}\right) =
$$
\begin{equation}  \label{f_Jmt}
=  f_0J^{1/8}\,\frac{\alpha + \ln\frac{m}{m_g}}{\alpha + \ln\frac{l_0\,J}{2m_g}}
  \frac{0.18}{R\,t} \exp\left(-\frac{{l_0}^2 J^2}{5\,R\,t}\right)
\end{equation}
For $r\sim 10$ pc $\alpha \sim 100$, therefore, the first fraction in Eq.(\ref{f_Jmt}) 
is close to unity.
The quantity $m_{max}$ in Eq.(\ref{rho_r}) is determined from the condition of 
zero denominator:
$m_{max} = 2\sigma_0r\sqrt{\ln [\sqrt{\pi}J/\sigma_0r]} \simeq 2\sigma_0r$
for $J>J_{min}=\sigma_0r/\sqrt{\pi}$.
However, approximation (\ref{f_Jmt}) is applicable only for low $m$ values and, 
in particular, the total integral $\int_0^{m_{max}} f(J,m,t)\,2m\,dm$ cannot be 
larger than $f_0J^{1/8}$ according to the normalization condition.
This gives the restriction $m \le m_0 = \sqrt{5\,R\,t}$, which becomes important
for $r>r_*\simeq 2$~pc. 
Thus, the internal integral in Eq.(\ref{rho_r}) is given by the expression
$$
\int\limits_{m_g}^{\makebox[1.6cm][l]{$\scriptstyle{\rm min}(m_{max},m_0)$}} 
\frac{2m\,dm\,r}{\sqrt{{m_{max}}^2-m^2}}
  \,\frac{0.18\,f_0J^{1/8}}{R\,t}\exp\left(-\frac{{l_0}^2 J^2}{5\,R\,t}\right) =
$$
$$
 = 2r\,\frac{0.18\,f_0J^{1/8}}{R\,t}\exp\left(-\frac{{l_0}^2 J^2}{5\,R\,t}\right)
  \times
$$
$$\times \left\{ \begin{array}{rcl}
  m_{max} = 2\sigma_0 r &,& r<r_* \\
  2\sigma_0 (r - \sqrt{r^2 - {r_*}^2}) &,& r>r_* \end{array} \right.
$$
Since $\d E/\d J = 2\pi/T(J) = 2{\sigma_0}^2/J$, we have
$$
\rho(r) = 16\pi\,{\sigma_0}^3\,
  {\rm min}\left[1,1-\sqrt{1-\left(\frac{r_*}{r}\right)^2}\right] \times
$$
$$
  \frac{0.18}{R\,t} \int_{J_{min}}^\infty dJ\,\frac{f_0J^{1/8}}{J} 
  \exp\left(-\frac{{l_0}^2 J^2}{5\,R\,t}\right) \simeq 
$$
$$
\simeq \frac{15\,{\sigma_0}^3}{R\,t}\,f_0\left(\frac{5\,R\,t}{{l_0}^2}\right)^{1/16}
  {\rm min}\left[1,1-\sqrt{1-\left(\frac{r_*}{r}\right)^2}\right] \simeq 
$$
$$
\simeq
10^3 \frac{M_\odot}{\mbox{pc}^3} \cdot \left(\frac{r}{10\mbox{ pc}}\right)^{-2}
$$
Thus, the estimate for the density agrees in order of magnitude with 
observations. In any case, the dark matter density does not exceed the star 
density, which is estimated as
$\rho_s(r)=1.2\cdot 10^6 (r/0.4\mbox{ pc})^\beta
M_\odot/\mbox{pc}^3$, where $\beta \simeq 1.5$ for $r<0.4$~pc 
and $\beta \simeq 2$ for $r>0.4$~pc \cite{genzel5}.
For a more accurate determination of the dark matter density profile, it is 
necessary to take into account the diffusion along the $J$ axis and a more exact 
expression for $f(J,m,t)$ in the limit $m\gg m_g$.
\section{Conclusions}
In this work, the diffusion of dark matter at the Galactic center has been 
considered. This diffusion is caused by scattering of dark matter particles on 
bulge stars and is considered in the $\{J, m, m_z\}$ phase space, where $J$ is 
the modified radial action.
Diffusion along the axis of the orbital angular momentum $m$ plays the main role 
in the absorption of dark matter. The presence of the black hole determines the 
boundary condition at $m=m_g$ that has different forms for small and large 
values of $J$. The dark matter flux $S(t)$ calculated using the modified 
boundary condition is much smaller than the value obtained in previous works, 
where this modification was disregarded (i.e., the absorbing boundary 
approximation was applied). The amount of dark matter absorbed by the black 
hole may at present compose a significant fraction of the black hole mass. The 
dark matter density profile induced by diffusion within the central region 10 
pc in size can be responsible for the observed $\gamma$ radiation from the Galactic 
center, which arises upon the annihilation of dark matter particles. 

We are grateful to A.S. Ilyin and V. I. Sirota for fruitful discussions. This 
work was supported by the Russian Foundation for Basic Research (project nos.
01-02-17829, 03-02-06745) and Landau Foundation (For\-schungs\-zent\-rum J\"ulich).

}
\end{document}